# Spectrally resolved far-field emission pattern of single photon emitters in MoS$_2$

Katja Barthelmi[1,2], Tomer Amit[3], Lukas Sigl[1], Mirco Troue[1,2], Thomas Klokkers[1], Anna Herrmann[1], Takashi Taniguchi[4], Kenji Watanabe[5], Jonathan Finley[1,2], Christoph Kastl[1,2], Sivan Refaely-Abramson[3,a)], and Alexander Holleitner[1,2,b)]

[1] Walter Schottky Institute and Physics Department, Technical University of Munich, Am Coulombwall 4a, 85748 Garching, Germany.

[2] Munich Center for Quantum Science and Technology (MCQST), Schellingstr. 4, 80799 Munich, Germany.

[3] Department of Molecular Chemistry and Materials Science, Weizmann Institute of Science, Rehovot, Israel.

[4] Research Center for Materials Nanoarchitectonics, National Institute for Materials Science, 1-1 Namiki, Tsukuba 305-0044, Japan.

[5] Research Center for Electronic and Optical Materials, National Institute for Materials Science, 1-1 Namiki, Tsukuba 305-0044, Japan.

a) Electronic addresses: sivan.refaely-abramson@weizmann.ac.il

b) Electronic addresses: holleitner@wsi.tum.de

**ABSTRACT**

We explore the optical dipole orientation of single photon emitters in monolayer MoS$_2$ as produced by a focused helium ion beam. The single photon emitters can be understood as single sulfur vacancies. The corresponding far-field luminescence spectra reveal several photoluminescence lines below the dominating luminescence of the exciton in MoS$_2$. These sub-bandgap emission lines were predicted by ab initio theory, but they have never been resolved in luminescence experiments because of their small amplitude. We reveal the lines by their dependence as a function of the photon energy and momentum as measured in the back focal plane of the optical circuitry. The agreement between theory and experiment suggests that the defect states interact strongly within the Brillouin zone.

*Introduction.* Single photon emitters (SPEs) are essential building blocks to realize photonic quantum technologies [1,2]. Particularly, SPEs in two-dimensional (2D) materials promise an improved scalability and a controlled generation due to their intrinsically reduced dimensions compared to bulk materials [3]. A scalable generation of SPEs is highly desirable and has been demonstrated in 2D materials, including hBN [4–8], $WS_2$ and $WSe_2$ [9–15], $MoS_2$ [16–18], $MoSe_2$ [19], and $MoTe_2$ [20,21]. Many of those approaches utilize strain through nanopillar arrays in combination with defect sites [7,10,20], while ion and electron beam irradiation techniques are suitable generation methods of defect-related quantum emitters as well [4,6,16]. Particularly, the reduced dimensionality of 2D materials is advantageous for the integration of SPEs into photonic platforms [2,22], such as waveguides and micro-resonators [23,24], and generally for their use in integrated quantum technologies [25]. In the case of $MoS_2$, a scalable way to generate quantum emitters was demonstrated by irradiating the material with a focused beam of He-ions [26]. The ion-based fabrication creates predominantly sulfur vacancies [27], and the corresponding sites can be positioned with a lateral accuracy below 10 nm [28] and in already assembled vdW heterostructures [29]. In turn, the defects can be integrated in gated structures, enabling the electrostatic control of single photon emission [29] and optoelectronic vertical tunnel circuits [30].

To design efficient SPE sources and integrate them into photonic circuits, a fundamental understanding of the transition matrix and the optical dipole moment of the underlying excitonic transitions is needed [25,31]. The transition matrix elements can be studied by measuring the far-field emission of the emitted light in the back focal plane (BFP) [32]. Such angle-resolved photoluminescence spectroscopy can give insights into the emission characteristics of bright and dark excitons [33,34]. In $WSe_2$, the dipole of the bright exciton is oriented in-plane, while it is oriented out-of-plane for the so-called gray exciton [35]. For strain-induced quantum emitters in $WSe_2$, it was reported that SPE luminescence lines undergo a transition from apparently in-plane to out-of-plane BFP patterns, if their luminescence wavelength is shorter or longer than about 750 nm [36]. Moreover, for SPEs in hBN, an efficient light collection by the help of microlenses was confirmed by BFP microscopy [37]. In $MoS_2$, the optical dipole orientation of the exciton was reported to be in-plane at room temperature [32]. The dipole orientation at low temperatures as well as the orientation of defect related emission in $MoS_2$ have not been investigated so far.

Here, we explore the far-field emission spectrum of SPEs in $MoS_2$ in the BFP and compare the obtained data to an analytical model as well as to ab initio results as computed from GW and Bethe-Salpeter equation approximations. The SPEs are generated by He-ion irradiation in $MoS_2$ and can be understood as single sulfur vacancies [27]. We realize a fiber-based BFP spectroscopy by collecting the emitted light in the BFP using a single-mode glass fiber as point detector in the Fourier plane. The detected light is guided to a spectrometer and analyzed spectrally. The demonstrated BFP spectroscopy enables us to reveal sub-band gap luminescence lines of the SPEs. So far, these lines have been predicted [38], but have not been resolved in luminescence due to their significantly low intensity. Their dependence upon photon energy and momentum suggests that the sub-band gap luminescence lines stem from defect-related exciton states which are spread across the whole Brillouin zone of the $MoS_2$ and whose optical dipole is oriented mostly in-plane. Our results provide fundamental insights into the emission characteristics of site-selectively created single photon emitters in monolayer $MoS_2$.

*Methods.* Figure 1(a) sketches the investigated sample geometry. A monolayer $MoS_2$ is encapsulated in a top (28.8 ± 0.1) nm and bottom layer (13.5 ± 0.2) nm of hBN on top of a Si substrate with a (295 ± 1) nm $SiO_2$ buffer layer. The assembled vdW heterostructure is irradiated by a focused He-ion beam to create optically active defects [16,17,26]. Figure 1(b) shows the photoluminescence (PL) spectrum of a typical SPE measured at a bath temperature of $T_{bath}$ = 1.7 K. The dominant emission line at $E_{photon}$ = 1.757 eV has an energy of ~190 meV below the neutral exciton ($X^0_{1s}$) of $MoS_2$, which is consistent with previous reports of a quantum emitter labeled as Q1 in $MoS_2$ [39], and it is attributed to a single sulfur vacancy [27,39]. At an energy of ~31 meV below the SPE, we detect a luminescence line with a much reduced amplitude, which has been interpreted as a local phonon mode (LPM) so far [16]. The normalized second-order correlation function $g^{(2)}(\tau)$ of Q1, as measured by a Hanbury-Brown-Twiss experiment, clearly shows anti-bunching with $g^{(2)}(0)$ = 0.23 ± 0.01, proving single photon emission of the defect-related PL [inset Fig. 1(b)].

To study the far-field emission pattern of the SPE by low-temperature BFP-imaging and spectroscopy, the sample is placed inside a closed cycle cryostat. The corresponding optical circuitry is sketched in Fig. 1(c). The photon emission from the sample is collected by a low-temperature objective with a numerical aperture of NA = 0.81. The far-field emission pattern can be observed in the BFP of the objective. We use two relay lenses in combination with an achromatic tube lens and a Bertrand lens outside of the cryostat to image the BFP on a CMOS-camera. A polarizer (polarizing beam splitter cube) allows distinguishing between s- and p-polarized light with respect to the axis of the polarizing optic. When using the camera, the spatial emission distribution is integrated spectrally, and an optical bandpass filter selects the investigated energy range. For the spectroscopy measurements, a single mode optical fiber is mounted on an *x*- and *y*- scanning axis within the BFP, instead of the camera. The fiber is connected to a spectrometer, which enables measuring a spectrally resolved far-field emission pattern of the SPEs.

*Experimental results.* Figure 2(a) shows the SPE-related photon emission in k-space with the underlying PL spectrum shown in Fig. 1(a). The energy range of the integrated PL is selected by a 711 nm optical bandpass filter with a width of 25 nm (cf. Fig. S1 in the Supplemental Material [40]). To improve the quality of the data, multiple images are recorded, while an additional $\lambda$/2-plate, inserted to the beam path, is rotated wrt. the polarizing optical elements [34]. This method softens the influence of disturbances in the image and results in the circular pattern visible in Fig. 2(a) and (d). The k-space image is spanned by the axes ($k_x/k_0$) and ($k_y/k_0$), which are perpendicular (*s*) and parallel (*p*) to the axis of the polarizing beam splitter cube. For clarity, the axes represent the orthogonal components of the in-plane photon wave vector $k_0 \cdot \sin(\theta)$, with $k_0 = E_{photon} / \hbar c$, with $\hbar$ the Planck's constant, *c* the speed of light, and $\theta$ the emission angle [34]. The experimental data are fitted by an analytical model following the source term model by Benisty et al. [41]. The latter calculates the dipole emission pattern depending on the dipole angle and emission energy, considering the propagation of the light through the layered materials forming the vdW heterostructure by utilizing transfer matrices. Figure 2(c) compares the corresponding s- (red) and p-polarized (blue) far-field emission pattern of the experimental data (points) with the fit (line) as a function of the polar emission angle $\theta$. The fit suggests an optical dipole angle of 82° ± 6° [Fig. 2(b)], with an in-plane contribution of (98 ± 2)% [34]. The black dashed line indicates the maximum collection angle ($\theta_{NA}$ = 54.1°) provided by the numerical aperture (NA) of the utilized objective. For comparing our results and the utilized setup with earlier studies, Fig. 2(d) shows data at an emission energy of the $X^0_{1s}$ in $MoS_2$ ($E_{photon}$ = 1.95 eV). Here, we utilize a Raman filter at 632 nm in combination with a 650 nm short pass filter (cf. Figs. S1 and S2 in the Supplemental Material [40]). Most prominently, for the $X^0_{1s}$, the center exhibits a small dip in k-space intensity, while for the SPE, the intensity of the emission pattern is maximum in the center of the image. Our fit reproduces the data of the $X^0_{1s}$ exciton in terms of a complete in-plane optical dipole orientation [Figs. 2(e) and (f)], as it is consistent with earlier studies at room temperature [32].

The above spectrally integrated BFP imaging technique has a limited capability to distinguish between different emission lines within the spectral bandwidth of the utilized optical bandpass filters. To improve the performance, we replace the camera with a single mode optical fiber that can be scanned across the BFP [cf. Fig. 1(c)]. At each position within the BFP, the collected photons are analyzed in a spectrometer. The gained spectral resolution is accompanied by a smaller collection efficiency, since the light is coupled into a single-mode fiber and then into the spectrometer. Therefore, the experiments presented below are performed without a polarizer to increase the number of detected photons. Results of a measurement with the polarizing optic in the beam path are shown in Fig. S3 of the Supplemental Material [40]. Figure 3(a) depicts the measured spectra along $k_y/k_0$ wrt. the center of the BFP ($k_x$ = 0) in a logarithmic scale. Without polarizer, the far-field emission pattern is radially symmetric, therefore the scanning axis in Fig. 3(a) is merely defined as $k/k_0$. The white dashed lines indicate the NA of the objective ($k/k_0$ = ± 0.81). Figure 3(b) shows the measured spectrum at $k/k_0$ = 0 [white dotted line in Fig. 3(a)] in a logarithmic scale. Besides the dominant emission line of Q1 at 1.757 eV, we observe multiple, less intense emission lines at lower energies. Two of them were already reported, such as the local phonon mode (LPM) [16] and the line at an energy around 1.7 eV categorized as Q2 [39]. The luminescence stemming from the colored regions are numerically integrated and analyzed in the following.

Figures 3(c)-(g) present the numerically integrated intensities as a function of polar emission angle for the five emission lines highlighted by the colored background correspondingly in Fig. 3(b). The model fit to each data set is displayed by gray lines, and the data are normalized by the number of integrated energy bins. The resulting degrees of in-plane polarization exhibit values of 100% for all emission lines analyzed. For comparison, Fig. 3(h) shows the polar dependence of the background signal centered around $E_{photon}$ = 1.92 eV, as will be discussed below.

*Theoretical results.* We compare our experimental findings to the theoretical excitonic picture obtained from an ab initio calculation based on many-body perturbation theory within the GW and Bethe-Salpeter equation (GW-BSE) approach. Figure 4(a) presents the correspondingly computed absorption spectrum for a linear in-plane light polarization. As we recently discussed [42,43], the defect-induced absorption shows up for energies below the strong $A$- ($X^0_{1s}$) and $B$-resonances. We highlight two main energy regions of this defect-induced absorption in orange and purple. For both regions, Figs. 4(b) and 4(c) show the normalized absorption as a function of the photon polarization angle. An out-of-plane (OOP) polarization angle of 0° corresponds to a purely in-plane polarization, as often used to examine excitons in layered TMDs, while an angle of ±90° describes a light polarization fully perpendicular to the monolayer plane. The computed photon polarization dependence on exciton energy is very weak. Figure 4(d) shows the out-of-plane exciton spin expectation values $<S|S_z|S>$ as a function of the exciton energy. Each exciton state is represented as a dot, with a size corresponding to the exciton oscillator strength for in-plane polarized light. This presentation demonstrates the distribution of spin expectation values based on the presence of the sulfur vacancy and the associated many-body nature of each exciton state, manifested in various contributing electron and hole states of varying localization. We observe exciton transitions with finite oscillator strength throughout the whole energy range despite having non-zero spin expectation values. In a naïve picture, the latter would resemble dark transitions, but our GW-BSE calculations demonstrate that this complex nature of the exciton states gives rise to an OOP interaction with light [e.g. Figs. 4(b) and (c)]. In this respect, the GW-BSE results demonstrate the many-body complexity beyond a simplistic picture; i.e. that the overall exciton spin is an outcome of mixed band transitions with various spin polarization [43].

*Discussion.* Generally speaking, the presented experimental luminescence results, the demonstrated source term model á Benisty [41] [lines in Fig. 2(c),(f) and Figs. 3(c)-(f)], and the GW-BSE are in very good agreement with each other. Moreover, with the reference measurement of the $X^0_{1S}$ exciton [Fig. 2(d)-(f)], our results consistently connect to reported measurements on the neutral exciton in $MoS_2$ at room temperature [32]. The autocorrelation measurement of $g^{(2)}(0)$ in the inset of Fig. 1(b) demonstrates that at least, the main resonance at $E_{photon}$ ~ 1.757 eV (baptized Q1 in ref. [39]) stems from a single exciton excitation. Unfortunately, autocorrelation measurements on the sub-bandgap luminescence lines are not possible in the current experimental configuration because of their very small luminescence amplitude [cf. inset of Fig. 1(b)] and long lifetime (~250 ns). Nevertheless, since the sub-bandgap emission lines with small amplitude [Fig. 3(b)] appear only alongside with the main resonance Q1 (cf. Fig. S1 in the Supplemental Material [40]), and since published photovoltage-measurements on comparable vertical tunnel devices also exhibit corresponding sub-bandgap absorption characteristics [30], we conclude that the sub-bandgap luminescence lines stem from the impact of single sulfur vacancies. We further observe that the sub-bandgap luminescence lines are not explained by the low-energy phonon-tail of Q1 (Fig. S4 in the Supplemental Material [40]) and the utilized instruments' background (Fig. S5 in the Supplemental Material [40]). When all previous arguments are combined, it follows that the sub-bandgap luminescence lines stem from the manifold exciton states and their light-matter interaction as predicted by the GW-BSE theory [cf. Fig. 4(d)]. Consistent with this interpretation, the energy dependences of the BFP emission characteristics of the sub-bandgap lines (Fig. 3) as well as of the calculated polarization-dependence of the absorption [e.g. Figs. 4(b),(c)] are equally very weak. However, the GW-BSE theory also predicts non-zero out-of-plane exciton spin expectation values $<S|S_z|S>$ with a finite oscillator strength over the whole range of sub-bandgap energies [Fig. 4(d)]. Experimentally, such contributions to the BFP-signal have their largest impact at large $k/k_0$ values; i.e. at the outer rim of the BFP where the signal is intrinsically low and the impact of (systematic) errors maximum. When we carefully analyze the data of Fig. 2(b), we indeed observe that the utilized source term model predicts an optical dipole angle of 82° ± 6° [Fig. 2(b)], with an in-plane contribution of 98 ± 2%, when the data at large $k/k_0$ are specifically taken into account (Fig. S5 in the Supplemental Material [40]) For the spectroscopic BFP data as in Fig. 3, the overall signal is reduced because of the utilization of the glass fiber, and even comparable to the given noise for large $k/k_0$ values (close to 0.81). In turn, for the BFP spectroscopy data as in Fig. 3, we cannot carve out the impact of the BFP data at high $k/k_0$ values as in the case of the spectrally integrated BFP (cf. Fig. S6 in the Supplemental Material [40]). Last but not least, the above reasoning suggests that the overall sub-bandgap luminescence background, sometimes referred to as "peak L" in earlier studies without and with hBN-encapsulation [16,44], has contributions from a manifold of the defect-related many-body exciton-states and that the so-called local phonon mode (LPM) as in Fig. 1(b) might be just one of many sub-band gap luminescence peaks.

*Conclusion.* We introduced a back focal plane (BFP) spectroscopy to reveal the orientation of the transition dipole moment of He-ion induced defects in monolayer $MoS_2$. By scanning across the BFP with a single mode fiber, we could resolve emission lines with low intensity below the energy of the neutral exciton. From our model fit, we found that all observed defect related luminescence is dominated by an in-plane transition dipole moment, which matches the results from GW-BSE computed absorption.

*Acknowledgements.* We gratefully acknowledge financial support by the Deutsche Forschungsgemeinschaft (DFG) via e-conversion – EXC 2089/1 – 390776260 and the Munich Center for Quantum Science and Technology (MCQST) – EXC 2111–390814868. C.K. and A.H. acknowledge support through TUM International Graduate School of Science and Engineering (IGSSE), and A.H and J.F. from the Munich Quantum Valley K6 and the One Munich Strategy Forum − EQAP. T.A. and S. R.-A. acknowledge support from the Minerva Foundation grant 7135421 and the European Research Council (ERC) Starting Grant (No. 101041159). T.A. acknowledges support from the Azrieli Graduate Fellows Program. Computational resources were provided by the ChemFarm local cluster at the Weizmann Institute of Science. K.W. and T.T. acknowledge support from the JSPS KAKENHI (Grant Numbers 21H05233 and 23H02052) and World Premier International Research Center Initiative (WPI), MEXT, Japan.

**Figures**

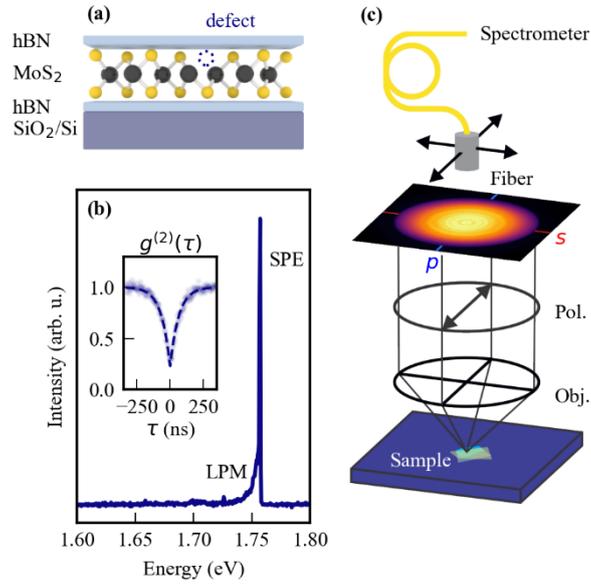

**FIG. 1.** Sample and measurement procedure. (a) Single photon emitters (SPEs) are generated in a van-der-Waals (vdW) heterostructure consisting of hBN-encapsulated $MoS_2$ on top of a Si/SiO$_2$ substrate. (b) PL spectrum of a SPE in $MoS_2$. A line with small amplitude is highlighted as local phonon mode (LPM). Measured at $T_{bath}$ = 1.7 K with an excitation of $E_{exc}$=1.962 eV and $P_{exc}$ = 500 nW. Inset shows second-order correlation function $g^{(2)}(\tau)$ measured at $T_{bath}$ = 10 K. (c) Sketch of the back focal plane (BFP) detection scheme with *p*- and *s*-polarization. The emission pattern is detected either by utilizing a camera or by scanning an optical fiber across the BFP.

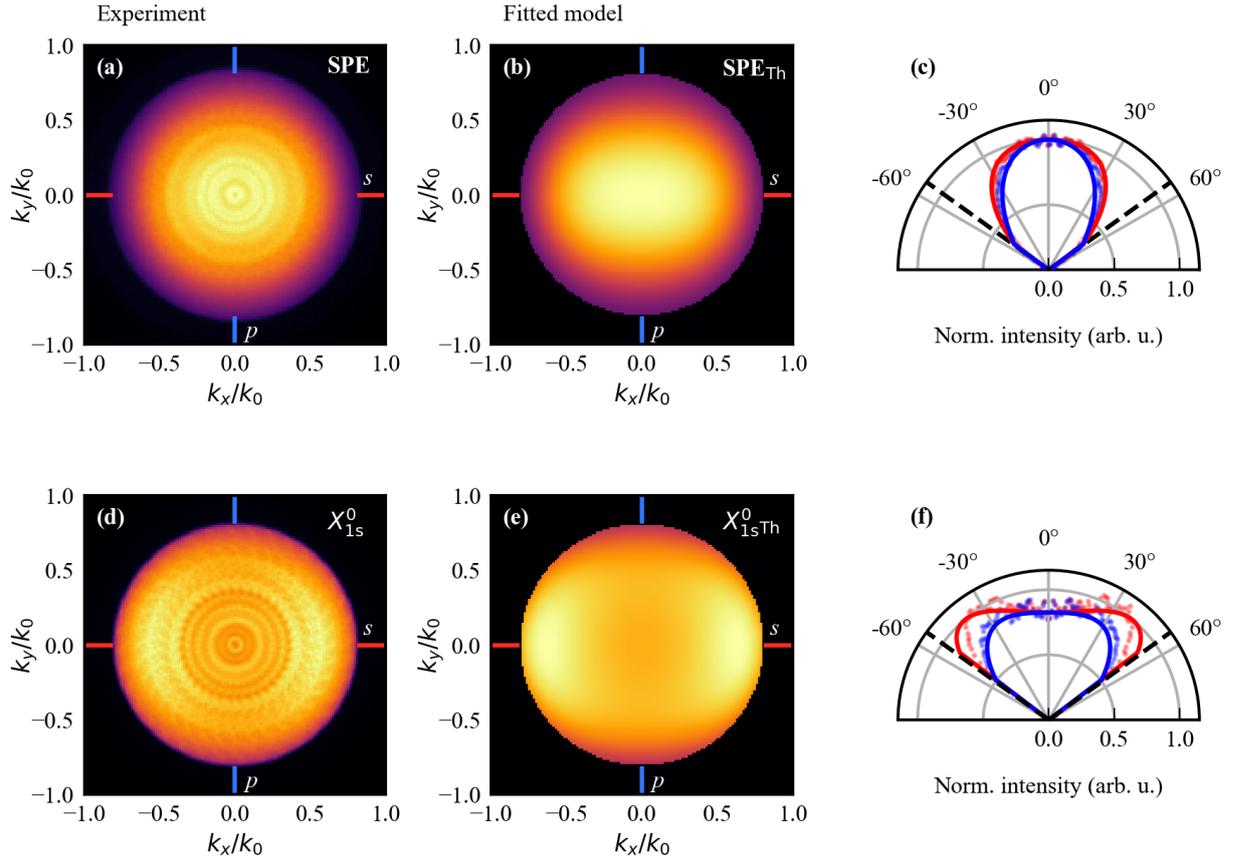

**FIG. 2.** (a) Spectrally integrated BFP image of a SPE at emission energy of $E_{photon} = 1.75$ eV shown in $k$-space. The $s$- ($p$-) polarization axis is marked in red (blue). (b) Model calculation in $k$-space (see text for details). (c) Line cuts through the $k$-space data (dots) for $s$- and $p$-polarization (red and blue) as well as fitting model (lines) displayed in polar coordinates and normalized to the fitting maximum. (d), (e), and (f) similar presentations for the PL stemming from the 1s exciton in MoS$_2$. Experimental parameters: $T_{bath} = 1.7$ K, $E_{exc} = 1.962$ eV, and $P_{exc} = 500$ nW.

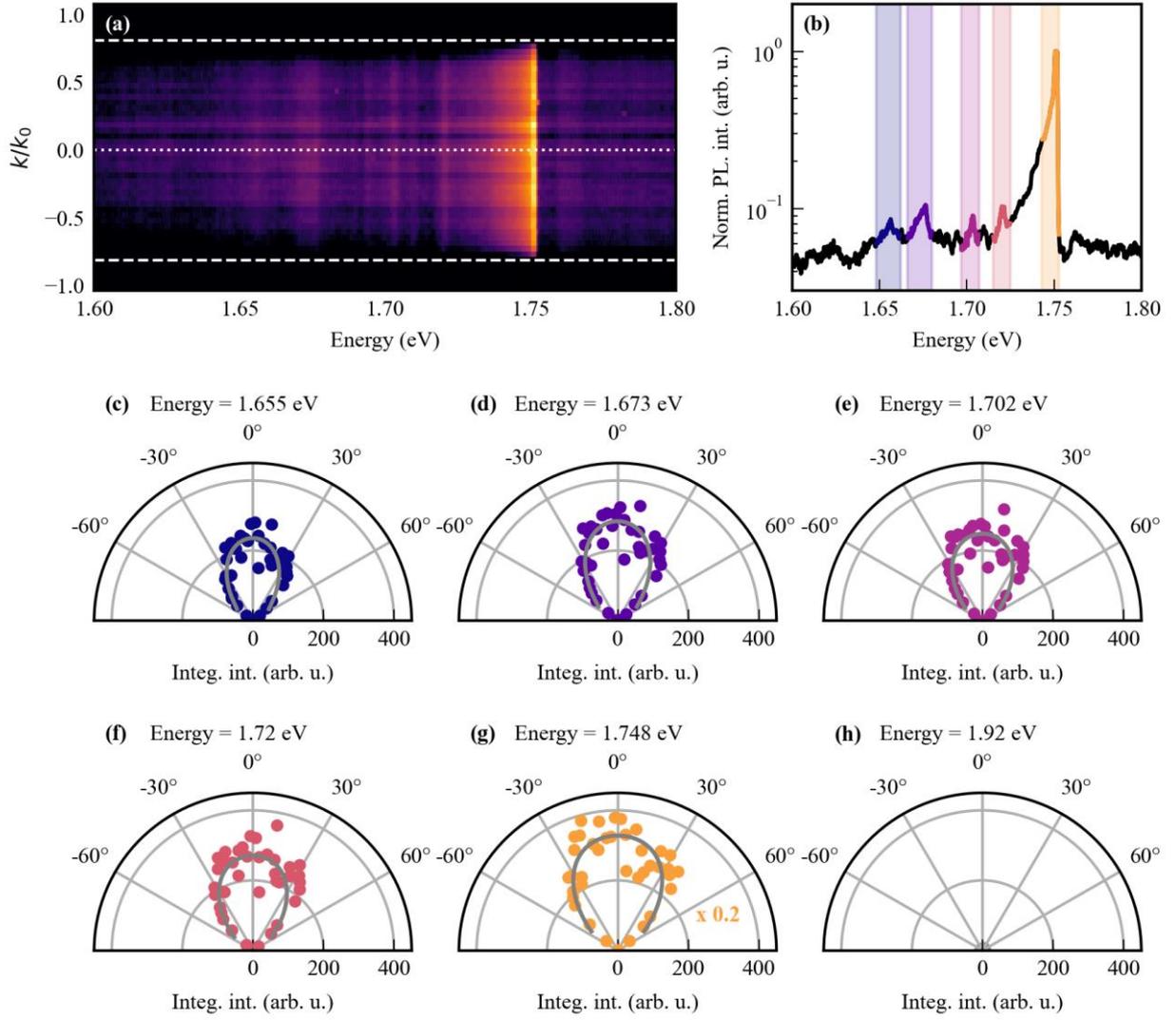

**FIG. 3.** Back focal plane spectroscopy. (a) Spectrally resolved PL of a SPE measured in the BFP of the utilized optical circuitry in dependence of the photon *k*-vector. Dashed lines indicate the NA of the objective. (b) Normalized PL spectrum at $k/k_0 = 0$ in logarithmic scale [cf. dotted line in (a)] Colors indicate energy intervals for sub-figures (c)-(g). (c)-(g) PL (dots) as highlighted in (b) presented in polar coordinates. The fits are depicted in grey, and the data is divided by the number of integrated energy bins. (h) Intensity of the background region, integrated from 1.91 to 1.93 eV. Experimental parameters: $T_{bath} = 1.7$ K, $E_{exc} = 1.962$ eV, and $P_{exc} = 500$ nW.

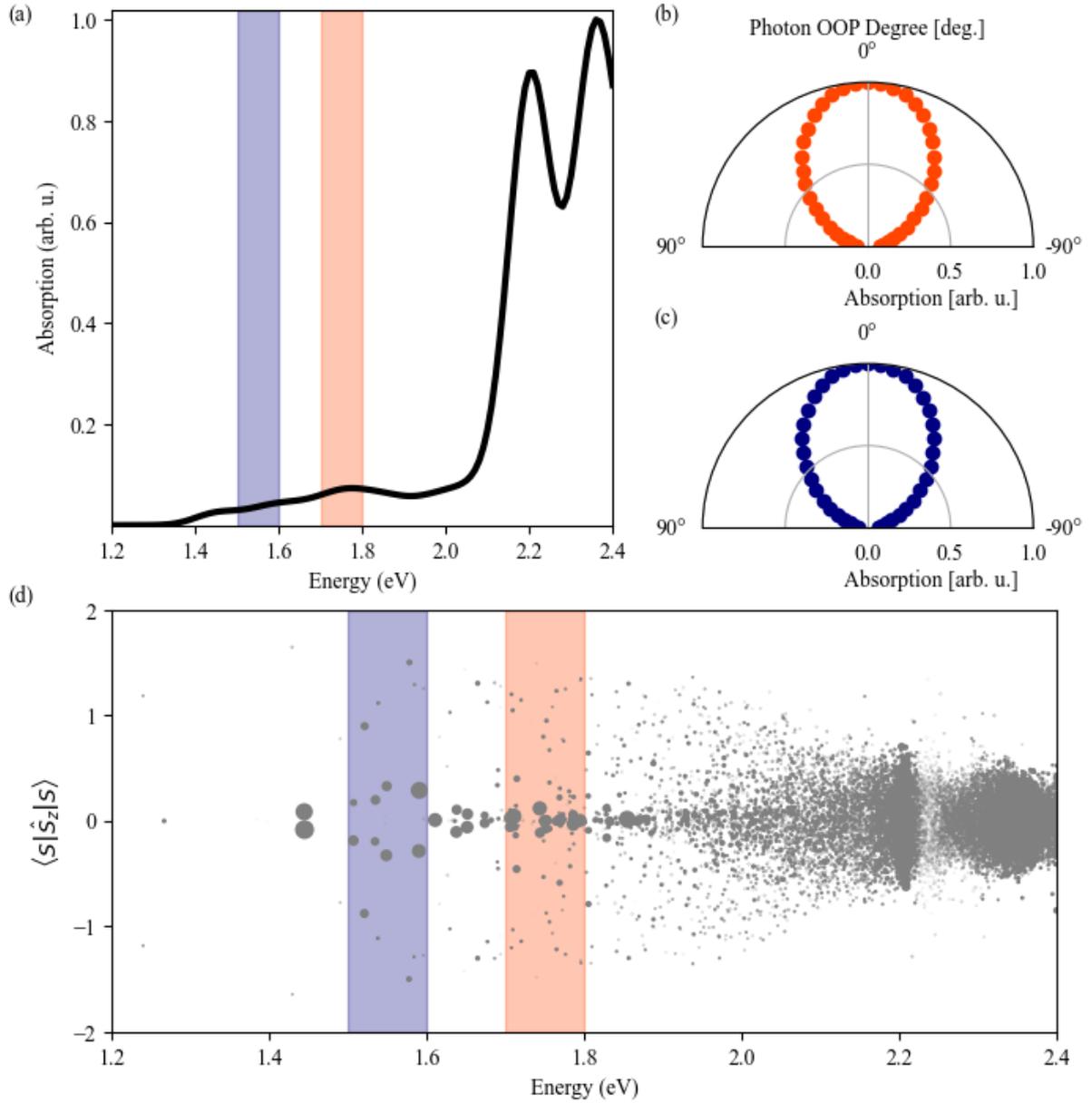

**FIG. 4.** Ab initio computations of a MoS$_2$ 5x5 supercell embedded with a single sulfur vacancy calculated with GW-BSE. (a) GW-BSE absorption spectrum featuring the *A*-like peak at 2.2 eV and the *B*-like peak slightly above. The spectrum shows a low-energy absorption region due to the sulfur vacancy. Two defect-related absorption energy regions are highlighted by purple and orange colors. (b) Normalized absorption for excitons at the orange-marked energy region, at different angles of the optical dipole with strongest absorption at zero degrees, corresponding to the absorption of an in-plane light polarization. (c) Same as (b) but for excitons at the purple-marked energy region. (d) Exciton spin expectation value along the out-of-plane direction at varying energies. The dot size corresponds to the excitonic oscillator strength for interacting with light polarized in-plane.

# Supplemental material for: Spectrally resolved far-field emission pattern of single photon emitters in MoS$_2$

Katja Barthelmi,[1,2] Tomer Amit,[3] Lukas Sigl,[1] Mirco Troue,[1,2] Thomas Klokkers,[1] Anna Herrmann,[1] Takashi Taniguchi,[4] Kenji Watanabe,[5] Jonathan Finley,[1,2] Christoph Kastl,[1,2] Sivan Refaely-Abramson,[3,a)] and Alexander Holleitner[1,2, b)]

[1] *Walter Schottky Institute and Physics Department, Technical University of Munich, Am Coulombwall 4a, 85748 Garching, Germany.*
[2] *Munich Center for Quantum Science and Technology (MCQST), Schellingstrasse 4, 80799 Munich, Germany.*
[3] *Department of Molecular Chemistry and Materials Science, Weizmann Institute of Science, Rehovot, Israel.*
[4] *Research Center for Materials Nanoarchitectonics, National Institute for Materials Science, 1-1 Namiki, Tsukuba 305-0044, Japan.*
[5] *Research Center for Electronic and Optical Materials, National Institute for Materials Science, 1-1 Namiki, Tsukuba 305-0044, Japan.*

**CONTENTS**

a) sivan.refaely-abramson@weizmann.ac.il
b) holleitner@wsi.tum.de

# I. SELECTION OF EMISSION RANGE

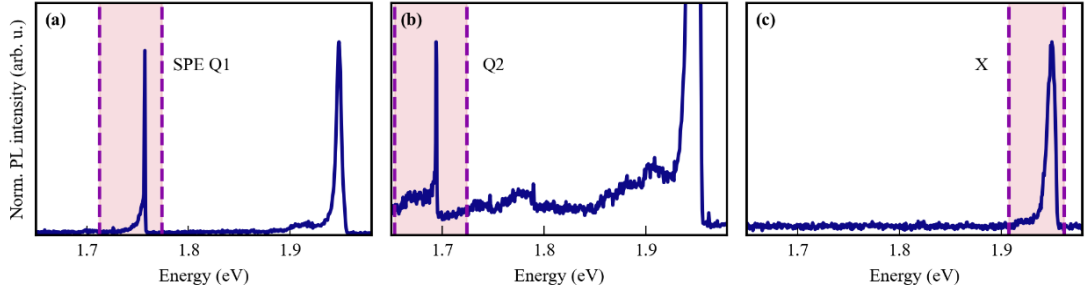

FIG. S1. **Filter regions for selection of specific emission ranges** shown in photoluminescence (PL) spectra for (a) single photon emitter (SPE) called Q1, (b) an emitter called Q2, and (c) the neutral $X^0_{1s}$ of $MoS_2$. PL spectra are recorded with $T_{bath}$ = 1.7 K, $P_{exc}$ = 500 nW and $E_{exc}$ = 1.962 eV. Colored areas indicate emission energy regions passed through the optical filters.

In the case of back focal plane images taken with the CMOS camera, the observed emission pattern is spectrally integrated over the energy range that gets to the detection path of the setup. To investigate the pattern for different emitters, the emission energy range is selected by optical filters. PL spectra of the emitters and the corresponding selected energy ranges are displayed in Fig. S1. The emission range for emitter Q1 at 1.75 eV can be selected by a 711 nm band pass filter (a), while for emitter Q2 at 1.69 eV a 735 nm band pass filter is used (b). The $MoS_2$ exciton emission is selected by a combination of a 632 nm razor edge Raman and a 650 nm short pass filter (c).

## II. EMITTER Q2 AT 1.69 EV

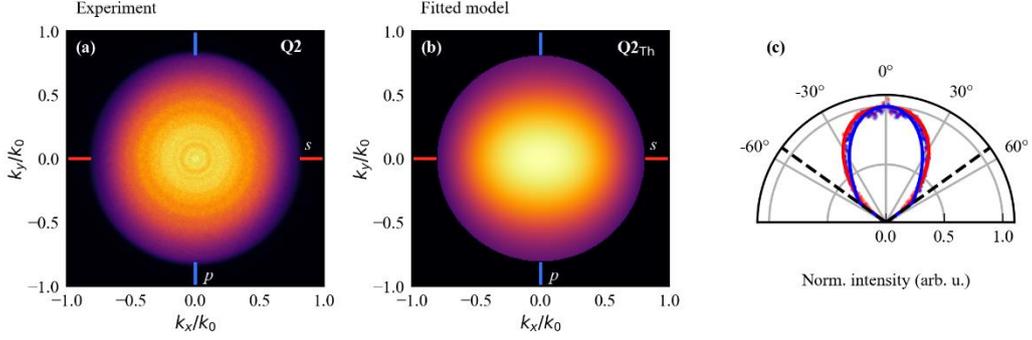

FIG. S2. **Spectrally integrated back focal plane (BFP) measurement** of a quantum emitter in $MoS_2$ with an emission energy of 1.69 eV (cf. e.g. Q2 in Fig. S1). Measured at $T_{bath}$ = 1.7 K and excited with $E_{exc}$ = 1.962 eV and $P_{exc}$ = 500 nW. (a) Back focal plane measurement shown as $k$-space image. The $s$- and $p$-polarization axes are marked by red ($s$) and blue ($p$) lines. (b) Emission pattern model fitted to the experimental data. (c) Line cuts through the $k$-space images for $s$- and $p$-polarization (red and blue) axes in the corresponding figures of the experimental data (dots) and the fitted model (line).

After the helium ion beam irradiation, additional emitters at an emission energy of 1.69 eV have been previously labeled as Q2 in previous experiments [1]. These emitters were also investigated with the first presented measurement technique of recording emission energy integrated back focal plane images. The experimental data is displayed in Fig. S2(a), and the model fitted to data in Fig. S2(b). The fit is compared to the data in Fig. S2(c) and results in an in-plane contribution of 100%. Comparing the emission pattern to the pattern of the emitter at 1.75 eV (main manuscript Fig. 2), it is noticeable that the pattern is less symmetric, with the emission along the $s$-axis being broader than along the $p$-axis.

## III. SPECTRALLY AND POLARIZATION-RESOLVED BACK FOCAL PLANE MEASUREMENT

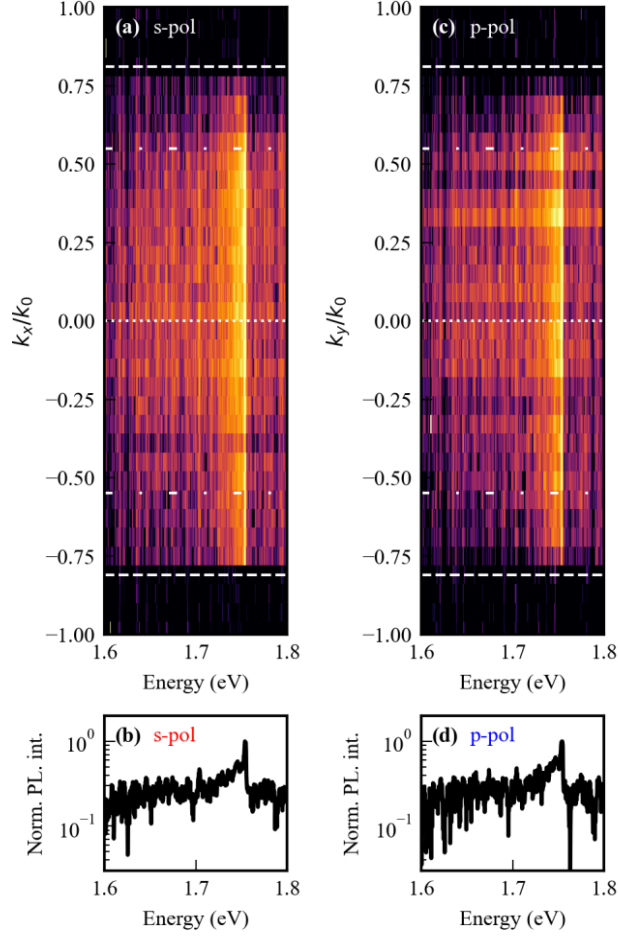

FIG. S3. **Spectrally and polarization-resolved BFP measurement** of a SPE in MoS$_2$ with an emission energy of 1.75 eV; measured at $T_{bath}$ = 1.7 K and excited with $E_{exc}$ = 1.962 eV and $P_{exc}$ = 500 nW. (a) Color map of the spectrally resolved PL in dependence of the $k$-vector along the $s$-polarization axis. Dashed lines indicate the NA of the objective. The dash-dotted line serves as a guide to the eye to compare emission patterns in (a) and (c). (b) PL spectrum at $k_x/k_0$ = 0 [dotted line in (a)]. (c) Color map of the spectrally resolved PL in dependence of the $k$-vector along the $p$-polarization axis. (d) PL spectrum at $k_y/k_0$ =0 [dotted line in (c)].

For investigating the dipole emission distribution in a spectrally resolved manner as in Fig. 3 of the main manuscript, the polarizing beam splitter was removed from the beam path because of the significantly reduced emission intensity. However, when it is installed, a polarization dependent measurement of the emission distribution in the BFP is possible by scanning across the two perpendicular axes, which is displayed in Fig. S3. Figure S3(a) shows the measured spectra along $k_x/k_0$ at the center ($k_y/k_0$ = 0) and therefore along the $s$-polarization axis. A PL spectrum at $k_x/k_0$ = 0 indicated by the dotted line is displayed in Fig. S3(b). Comparing the intensities of the PL spectra with the spectrum taken without a polarizer in Fig. 3(b) of the main manuscript, a reduced intensity of the emission can be observed. Therefore, the analysis of emission contributions with lower intensity was performed without a polarizing optic in the beam path. However, comparing $s$- and $p$-polarization axes a slight difference is noticeable with the emission being dimmer on the $p$-polarization axis. The observed asymmetry fits the emission model, which shows a broader distribution on the $s$- than on the $p$-axis.

## IV. FURTHER ANALYSIS OF THE SPECTRALLY RESOLVED BACK FOCAL PLANE MEASUREMENT

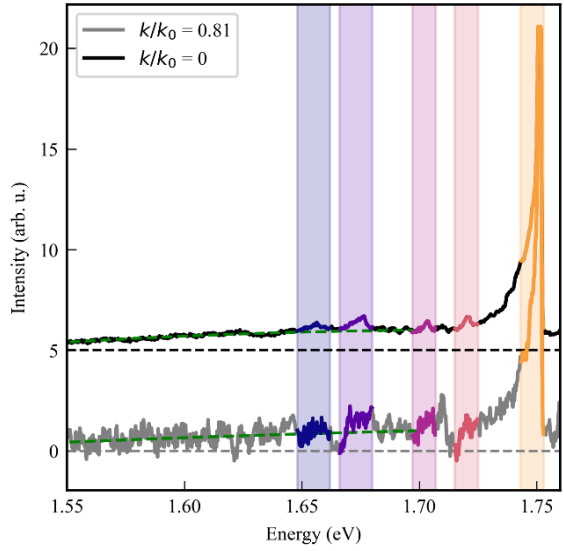

FIG. S4. **PL spectra at two maximally different $k$-values in the BFP.** Black (gray): PL spectrum at $k/k_0 = 0$ ($k/k_0 = 0.81$). Both spectra are normalized to the intensity of the broad underlying peak fitted by a polynomial (dashed green line). Zero-lines for both spectra are marked by dashed lines. Integration areas for the peaks are highlighted according to Figs. 3 and S5.

The comparison between a PL spectrum at the center of the BFP and the edge of the BFP with high $k$-values reveals that the peaks at smaller energy and with smaller amplitude are indeed photons stemming from the sample. For the comparison, the broad underlying peak in each of both spectra are fitted by a polynomial line, and their maximal intensities are used to scale the PL spectra. Through this step, it becomes apparent that the smaller peaks, which are close to the noise level in the spectrum at the edge of the BFP, rise clearly above noise level in the spectrum at the center of the BFP. Additionally, the underlying peak matches the energy of the multitude of exciton energies shown in Fig. 4(d), which appear to become visible in the spectrally resolved BFP measurements in contrast to normal PL.

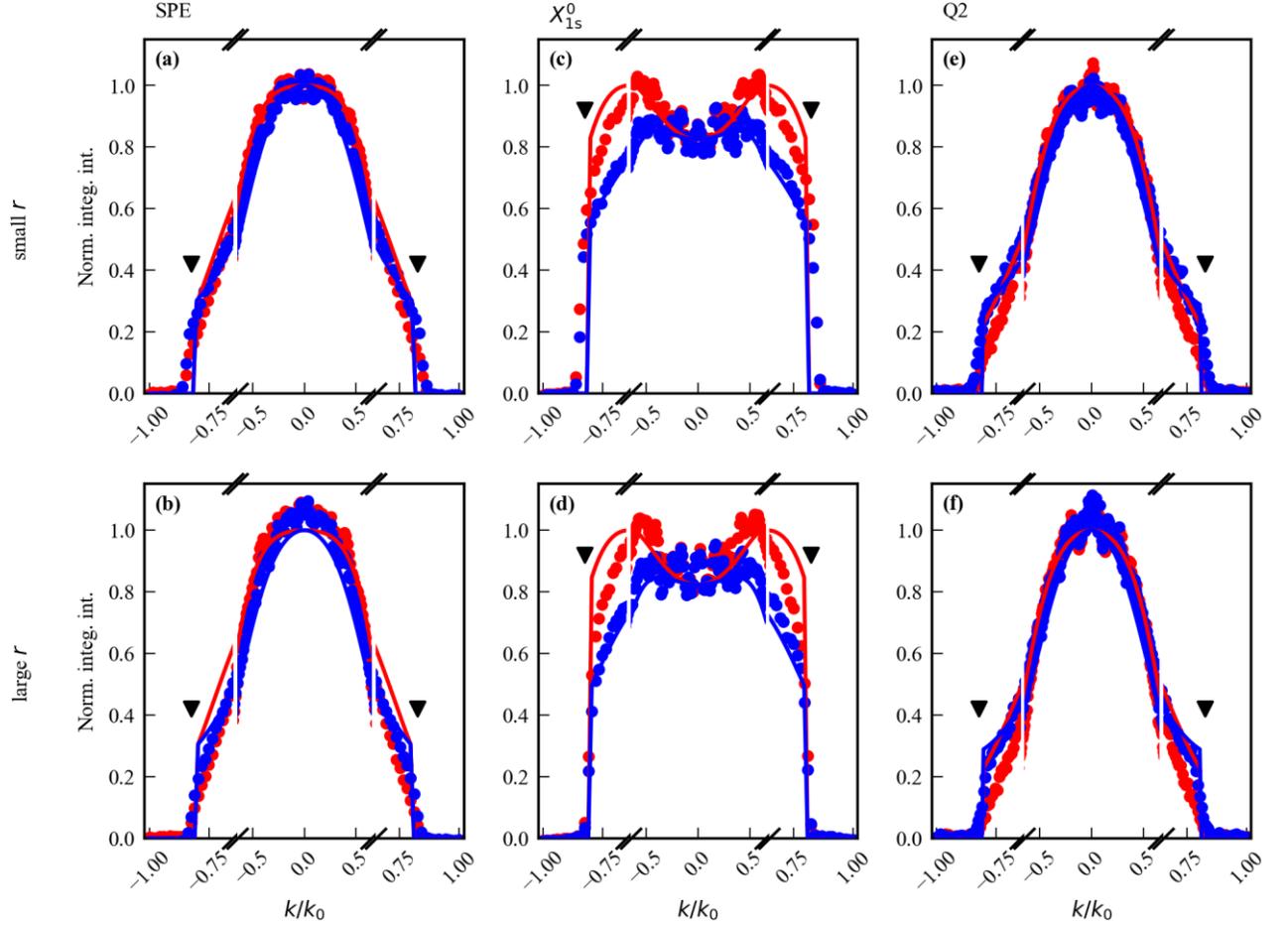

FIG. S5. **Influence of the pattern radius on the fit results.** (a) and (b) show the emission of the SPE, (c) and (d) the neutral $X^0_{1s}$ exciton and (e) and (f) the emission of Q2; along the *s*- and *p*-polarization axes (red and blue) fitted by a model (lines). Black triangles mark regions with high *k*-values. For (a), (c) and (e), the data was processed with a smaller pattern radius, while for (b), (d) and (f), a larger radius was used.

Figure S5 shows the cuts through the BFP images in Fig. 2 of the main manuscript and Fig. S2 in dependence of the *k*-vector instead of the emission angle. In combination with a different axis scale on the outer regions of the BFP, with larger *k*-values, this presentation allows a closer look at the sharp edge close to $k/k_0 = 0.81$. In processing the data, a pattern radius *r* is chosen, which has to fit to the edge that is caused by the NA of the objective. When fitting the emission of the SPE in (a) and (b), the fit with a larger pattern radius (of $r = 268.8\ \mu$m) does not match the data equally well. However, for the exciton in (c) and (d) the smaller pattern radius ($r = 259.2\ \mu$m) results in an emission pattern that is too small to match the sharp edge. Therefore, for each measurement the best pattern radius is chosen. For the emitter Q2 in (e) and (f) there is no strong difference. In this case, the smaller radius was chosen, since it creates a smaller $r^2$ value.

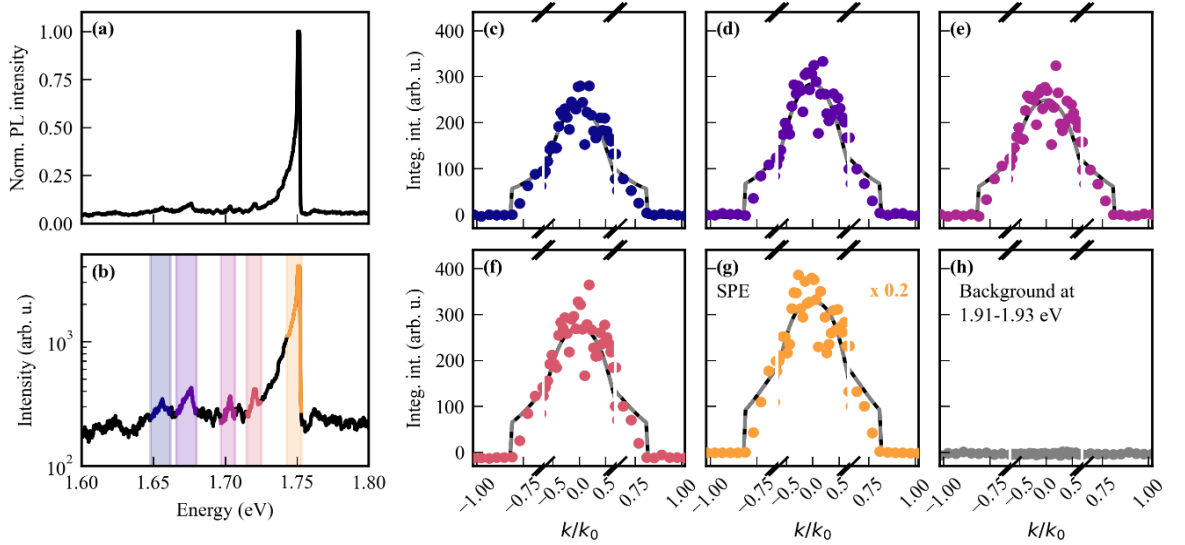

FIG. S6. **Analysis of the spectrally resolved BFP measurement** shown in Fig. 3 of the main manuscript, here displayed in $k$-space. (a,b) Normalized PL spectrum at $k/k_0 = 0$ in a linear scale (a) and logarithmic scale (b). Colors indicate energy intervals as for the intensities in sub-figures (c)-(g). (c)-(g) Emission pattern (dots) of the PL peaks and the background (gray) shown in (b) divided by integrated energy bins. The fit to the data is displayed in the black line. For comparison, the model of an 80° oriented dipole is displayed in the grey dashed line.

In the $k$-space dependent presentation of the BFP spectroscopy measurement in Fig. S6, especially the non-angular dependent emission intensity of the background (h) becomes more visible. Like Fig. S4, the scale on the $x$-axis [in (c) to (g)] changes for larger $k$-values to allow a closer look on this region. Due to the reduced sampling rate of the PL spectra at each position compared to the data points in a BFP image, no sharp edge at the NA of the objective ($k/k_0 = 0.81$) is visible, which might result in a less accurate fit. For comparison, a model with a dipole angular of 80° is plotted in the grey dashed line. It follows the same shape as the model fit with a dipole angle of 90°. The similarity of the emission models for mostly in-plane oriented dipole angles inside the observable region given by the NA of the objective could explain the slightly different measurement results from the BFP imaging and the BFP spectroscopy for the SPE with an energy of 1.75 eV.